\pdfoutput=0
\documentclass[11pt]{article}
\usepackage{epsfig}
\usepackage{amsfonts}
\usepackage{amsmath}
\usepackage{bm,bbm}
\usepackage{cite}
\allowdisplaybreaks[1]

\newcommand{\la}[1]{\label{#1}}
\newcommand{\be}{\begin{equation}}
\newcommand{\ee}{\end{equation}}
\newcommand{\ba}{\begin{eqnarray}}
\newcommand{\ea}{\end{eqnarray}}
\newcommand{\fig}{Fig.~}

\newcommand{\eq}{Eq.~}
\newcommand{\eqs}{Eqs.~}
\newcommand{\se}{Sec.~}

\newcommand{\nr}[1]{(\ref{#1})}

\newcommand{\nn}{\nonumber \\}

\newcommand{\tfr}[2]{{\textstyle \frac{#1}{#2}\,}}
\renewcommand{\eq}{eq.~}
\renewcommand{\eqs}{eqs.~}
\renewcommand{\se}{sec.~}

\renewcommand{\fig}{fig.~}

\newcommand{\varh}{\mathrm{h}}
\newcommand{\now}{{\mbox{\tiny\rm{0}}}}
\newcommand{\anow}{a_\now}
\newcommand{\fnow}{f_\now}
\newcommand{\snow}{s_\now}
\newcommand{\Hnow}{H_\now}
\newcommand{\Tnow}{T_\now}

\newcommand{\gw}{\rmi{gw}}
\newcommand{\igw}{\rmii{gw}}
\newcommand{\bbn}{\rmi{bbn}}
\newcommand{\Tstar}{T_{*}}

\newcommand{\rad}{\rmi{dv}} 
\newcommand{\irad}{\rmii{dv}} 
\newcommand{\dr}{\rmi{d}} 
\newcommand{\vr}{\rmi{v}} 
\newcommand{\idr}{\rmii{d}} 
\newcommand{\rd}{-} 
\newcommand{\dd}{+} 

\newcommand{\rmO}{{\mathcal{O}}}

\def\lsi{\raise0.3ex\hbox{$<$\kern-0.75em\raise-1.1ex\hbox{$\sim$}}}
\def\gsi{\raise0.3ex\hbox{$>$\kern-0.75em\raise-1.1ex\hbox{$\sim$}}}
\newcommand{\lsim}{\mathop{\lsi}}

\newcommand{\rmi}[1]{{\mbox{\scriptsize #1}}}
\newcommand{\rmii}[1]{{\mbox{\tiny\rm{#1}}}}

\newcommand{\Tint}[1]{{\hbox{$\sum$}\!\!\!\!\!\!\!\int\,}_{\!\!\!\!\raise-0.9ex\hbox{$\scriptstyle{#1}$}}}
\newcommand{\Tinti}[1]{{{\Sigma}\!\!\!\!\raise0.3ex\hbox{$\int$}_\rmii{${#1}$}}}
\newcommand{\bi}{\begin{itemize}}
\newcommand{\ei}{\end{itemize}}
\newcommand{\hide}[1]{ }

\newcommand{\deltabar}{\raise-0.02em\hbox{$\bar{}$}\hspace*{-0.8mm}{\delta}}
\newcommand{\ddeltabar}{\raise-0.18em\hbox{$\bar{}$}\hspace*{-0.8mm}{\delta}}
\newcommand{\B}{\rmii{$B$}}
\renewcommand{\H}{\rmii{$H$}}

\newcommand{\mpl}{m_\rmii{pl}} 

\usepackage{color}

\makeatletter \@addtoreset{equation}{section} \makeatother
\renewcommand{\theequation}{\arabic{section}.\arabic{equation}}
\makeatletter
\renewcommand\section{\@startsection {section}{1}{\z@}%
                                   {-5.5ex \@plus -1ex \@minus -.2ex}
                                   {2.3ex \@plus.2ex}%
                                   {\normalfont\large\bfseries}}
\renewcommand\subsection{\@startsection{subsection}{2}{\z@}%
                                     {-3.25ex\@plus -1ex \@minus -.2ex}%
                                     {1.5ex \@plus .2ex}%
                                     {\normalfont\normalsize\bfseries}}
\renewcommand\thesection {\@arabic\c@section}
\renewcommand\thesubsection   {\thesection.\@arabic\c@subsection}
\renewcommand{\@seccntformat}[1]{%
\csname the#1\endcsname.\hspace{1.0em}}
\makeatother

\begin{document}

\flushbottom

\begin{titlepage}

\begin{flushright}
February 2024
\end{flushright}
\begin{centering}
\vfill

{\Large{\bf
 Update on gravitational wave signals from
 \\[3mm] 
 post-inflationary phase transitions 
}} 

\vspace{0.8cm}

H.~Kolesova${}^\rmi{a}_{ }$ and
M.~Laine${}^\rmi{b}_{ }$

\vspace{0.8cm}

${}^\rmi{a}_{ }${\em
Faculty of Science and Technology, 
University of Stavanger, \\ 
4036 Stavanger, Norway \\}

\vspace{0.3cm}

${}^\rmi{b}_{ }${\em
AEC, 
Institute for Theoretical Physics, 
University of Bern, \\ 
Sidlerstrasse 5, CH-3012 Bern, Switzerland \\}

\vspace*{0.8cm}

\mbox{\bf Abstract}
 
\end{centering}

\vspace*{0.3cm}
 
\noindent
In view of recent interest in high-frequency detectors, 
broad features of 
gravitational wave signals from phase transitions taking place soon
after inflation are summarized.
The influence of the matter domination era
that follows the slow-roll stage is quantified
in terms of two equilibration rates.
Turning to the highest-frequency part of the spectrum, 
we show how it is constrained by the fact
that the bubble distance scale must exceed the mean free path.

\vfill


\end{titlepage}


%
\section{Introduction}
\la{se:intro}

A cosmological 
first-order phase transition may produce 
a gravitational wave signal~\cite{old1,old2,old3}. 
The signal is expected to be peaked, 
with the peak frequency proportional to the phase transition 
temperature, $\Tstar^{ }$. 
In particular, phase transitions at the QCD scale, 
$\Tstar^{ }\sim 100$~MeV, could yield a signal for
the high end of the  
pulsar timing array observation window
($\fnow^{ } \sim 10^{-8}_{ }...10^{-6}_{ }$~Hz); 
those at the weak scale, with 
$\Tstar^{ }\sim 100$~GeV, would match the LISA frequency range
($\fnow^{ } \sim 10^{-4}_{ }...10^{-1}_{ }$~Hz); 
whereas exotic phase transitions with
$\Tstar^{ }\sim 1 ... 100$~TeV could leave an imprint to be 
discovered by 
DECIGO ($\fnow^{ } \sim 10^{-1}_{ }... 10^{1}_{ }$~Hz) 
or the Einstein Telescope 
($\fnow^{ } \sim 10^{1}_{ }... 10^{3}_{ }$~Hz)~\cite{old4}. 
Phase transitions at much higher scales, 
related perhaps to GUTs, might in turn be probed
at the GHz...THz range that has become
of recent interest~\cite{uhf}. 

The signal from high-scale phase transitions has been 
estimated in ref.~\cite{uhf}. 
The purpose of the current study, which was influenced by 
recent model investigations~\cite{Tmax,new1}, 
is to derive an upper bound for a phase-transition-induced gravitational
wave energy density in the GHz...THz range. 
We establish its peak frequency, 
and elaborate on the effect of
a matter domination era that follows slow-roll inflation
(cf.,\ e.g.,\ 
refs.~\cite{matter1,matter2,matter3}), 
if the relevant equilibration rates
are small compared with the Hubble rate. 
The main differences with respect to ref.~\cite{uhf} will be 
recapitulated in the conclusions (cf.\ \se\ref{se:concl}). 
Before that, we review the main features of the background evolution
in the reheating epoch (cf.\ \se\ref{se:bg}), 
broad features of the gravitational wave signal originating
from phase transitions (cf.\ \se\ref{se:gw}), and the maximal
strain that this physics can lead to (cf.\ \se\ref{se:max}).

%
\section{Background evolution}
\la{se:bg}

We have in mind a usual single-field inflationary scenario. 
As the slow-roll stage ends, the inflaton starts
oscillating and the temperature reaches a maximal value.
There is normally a period in which inflaton oscillations 
dominate the overall energy density, 
before the energy density of a thermal plasma takes over. 

We assume the matter content during this time 
to reside in three components:
\bi

\item {\bf inflaton}, $\varphi$, with energy density $e^{ }_\varphi$.

\item {\bf ``dark sector''}, 
defined as the one undergoing a phase transition. 
The low-temperature phase of the dark sector is assumed massive
(i.e.\ non-relativistic), due to confinement or 
Higgs mechanism, so it does not exert pressure. 
Its energy density is denoted by $e^{ }_\dr$.

\item {\bf ``visible sector''}, 
constituted of Standard Model like particles, 
which are effectively massless
during the epoch under consideration.
Its energy density is denoted by $e^{ }_\vr$. 

\ei
Adding sectors together, 
we denote $e^{ }_{ij} \equiv e^{ }_i + e^{ }_j$, 
notably
$
 e^{ }_\rad \equiv e^{ }_\dr + e^{ }_\vr
$.
The overall energy density is then 
$
 e \equiv e^{ }_{\varphi\rad}
$, 
with the corresponding Hubble rate given by
\be
 H^2 = \frac{8\pi e}{3 \mpl^2}
 \;, \quad
 \mpl^{ } = 1.22091 \times 10^{19}_{ } \; \mbox{GeV}
 \;. 
\ee
In general we parametrize moments in the reheating epoch
by the value of $H$. Motivated by Planck data~\cite{planck}, 
we also conservatively set 
$
 H \le H^{ }_\rmi{max} \equiv 10^{-5}_{ }\mpl^{ }
$.

We model the dynamics of this coupled system by equations of the type
\be
 \dot{e}^{ }_i + 3 H (e^{ }_i + p^{ }_i) 
 \;\simeq\;
 - {\textstyle \sum^{ }_j }\,\Gamma^{ }_{ij} 
 (e^{ }_j - e^{ }_{j,\rmi{eq}})
 \;. \la{general}
\ee
Summing over $i$, 
the Friedmann equation needs to be recovered, implying
$\sum^{ }_i \Gamma^{ }_{ij} = 0$.
The equilibrium values, $e^{ }_{j,\rmi{eq}}$, represent the fixed
point that would be attained if the system had infinite time. In 
order to simplify the setup, we assume that the visible sector 
is effectively thermalized, $e^{ }_{\vr} \approx e^{ }_{\vr,\rmi{eq}}$. 
Only two rates are assumed to have a large influence:\footnote{%
  In this setup the dark sector serves as 
  a mediator between the inflaton and the visible sector, however 
  we hope that the results model qualitatively also situations
  in which more than two rate coefficients play a role. 
 } 
\bi

\item {\bf inflaton interaction rate}, 
$\Upsilon \equiv \Gamma^{ }_{\varphi\varphi} = - \Gamma^{ }_{\dr\varphi}$.
In general $\Upsilon$ would be a function of the dark sector temperature, 
but we assume that the part originating from vacuum decays dominates. 
A possible temperature dependence can subsequently be mimicked by 
interpolating between various constant values.

\item {\bf dark sector interaction rate},
$\Gamma \equiv \Gamma^{ }_{\dr\dr} = - \Gamma^{ }_{\vr\dr}$. 

\ei
Assuming also an initial state 
with $e^{ }_\varphi \gg e^{ }_{\varphi,\rmi{eq}}$
and
$e^{ }_{\dr} \gg e^{ }_{\dr,\rmi{eq}}$, 
we are then faced with 
\ba
 \dot{e}^{ }_\varphi + 3 H (e^{ }_\varphi + p^{ }_\varphi)
 & \simeq & -\, \Upsilon\, e^{ }_\varphi
 \;, \la{eom_varphi}  \\
 \dot{e}^{ }_\dr + 3 H (e^{ }_\dr + p^{ }_\dr )
 & \simeq & +\, \Upsilon\, e^{ }_\varphi - \Gamma\, e^{ }_\dr
 \;, \la{eom_dr} \\ 
 \dot{e}^{ }_\vr + 3 H (e^{ }_\vr + p^{ }_\vr )
 & \simeq & +\, \Gamma\, e^{ }_\dr 
 \;. \la{eom_vr}
\ea
We define
\be
 H^{ }_\dd \; \equiv \; \max\{ \Upsilon,\Gamma \}
 \;, \quad 
 H^{ }_\rd \; \equiv \; \min\{ \Upsilon,\Gamma \}
 \;, \la{H_dd_rd}
\ee
with the corresponding scale factors denoted by 
$
 a^{ }_\dd
$
and 
$
 a^{ }_\rd
$, 
respectively. 
We now consider the solution 
of \eqs\nr{eom_varphi}--\nr{eom_vr} in various regimes,
making use of the labelling in \fig\ref{fig:cartoon}.  

%
\paragraph{Domain~(1).} 
During the reheating period, the inflaton is oscillating faster
than the Hubble rate, satisfying then 
$
 \ddot{\bar\varphi} + \partial^{ }_{\bar\varphi} V \simeq 0
$.
After multiplication with $\dot{\bar\varphi}$, this can be 
integrated into 
$
 \frac{1}{2}\dot{\bar\varphi}^2 + V \simeq 
$ constant. By the virial theorem, 
the two terms contribute equally on average,   
implying that $p^{ }_\varphi \simeq 0$, 
$e^{ }_\varphi \simeq \dot{\bar\varphi}^2$.
The regime in which these equations apply corresponds to  
a matter domination era. 
Recalling 
$
 e^{ }_\varphi \gg \max\{ e^{ }_\dr,e^{ }_\vr \}
$, 
so that $H^2 = 8\pi e^{ }_\varphi/(3\mpl^2)$, 
and $\Upsilon \ll H$, 
whereby the right-hand side can be omitted, 
\eq\nr{eom_varphi} yields 
$
 8\pi e^{ }_\varphi/(3\mpl^2)  \simeq [{2}/(3t)]^2_{ } 
$.
This in turn implies 
$
 H / H^{ }_m \simeq t^{ }_m/t
$, 
$
 (e^{ }_\varphi / e^{ }_{\varphi,m})
 \simeq 
 (t^{ }_m/t)^2
$, 
$
 a/a^{ }_m \simeq (t / t^{ }_m)^{2/3}_{ }
$, 
and 
$
 H^{ }_m t^{ }_m \simeq 2/3
$, 
where $t^{ }_m$ denotes the time at which matter domination starts.

%
\begin{figure}[t]

\centerline{
     \epsfysize=7.0cm\epsfbox{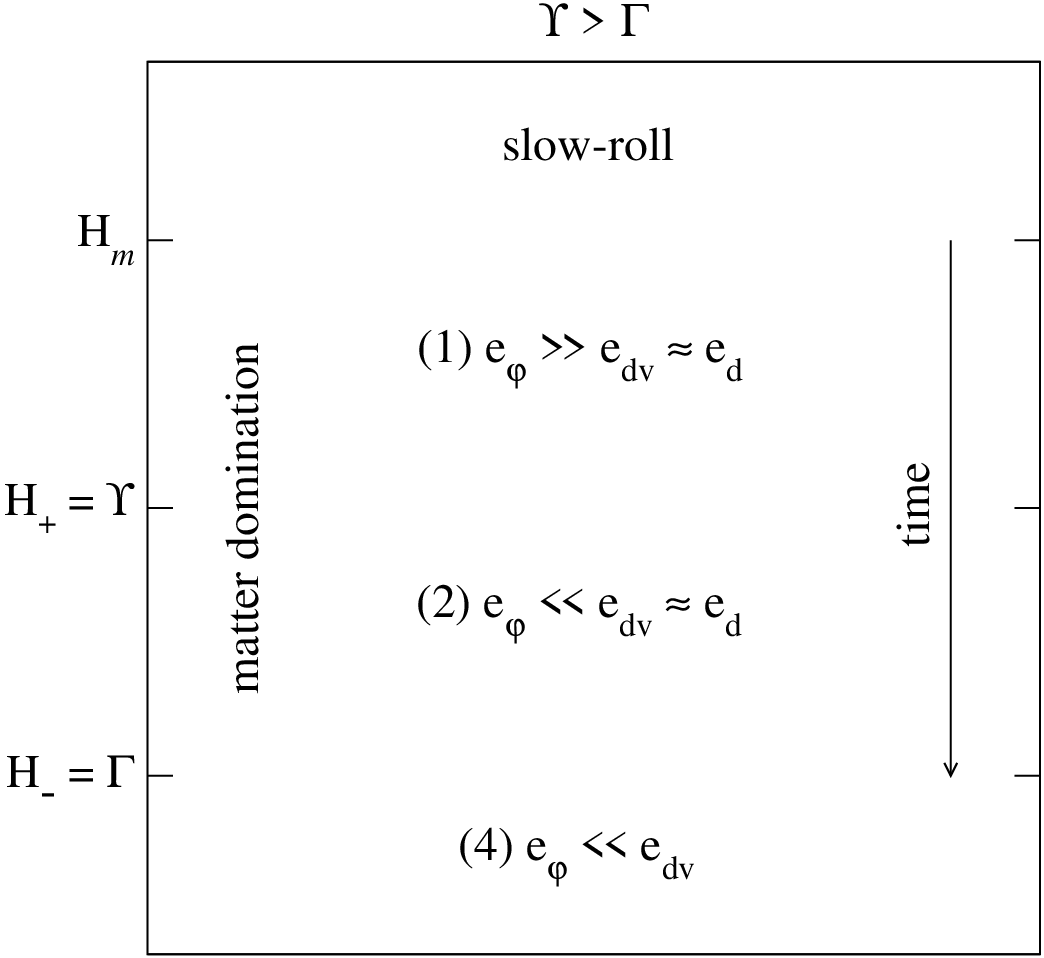}
  ~~~\epsfysize=7.0cm\epsfbox{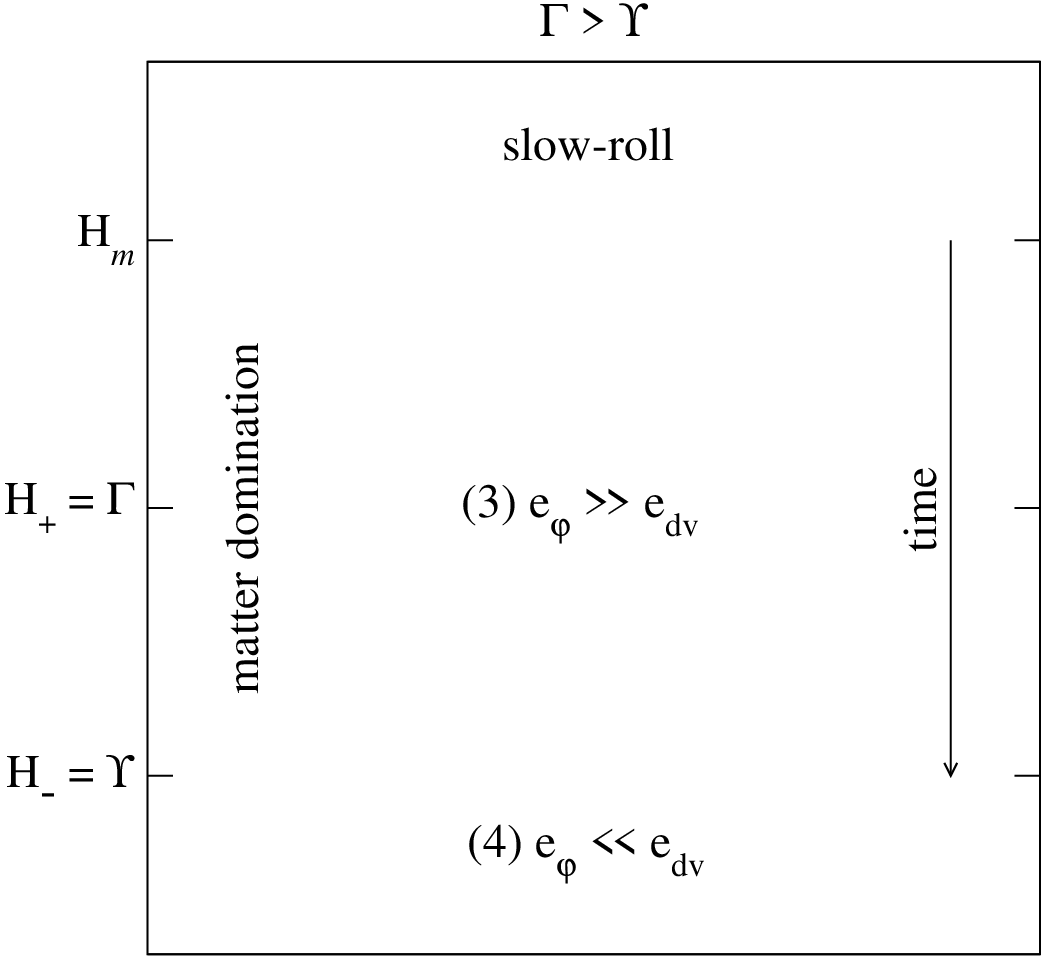}
}

\vspace*{1mm}

\caption[a]{\small 
 Cartoons of the key epochs of the reheating period.
 Here $H^{ }_m$ denotes
 the moment 
 at which the slow-roll expansion ends and the oscillation
 (i.e.\ matter domination)
 period begins, 
 and~$\Upsilon$ and~$\Gamma$ two rates that characterize the 
 dynamics (cf.\ \eqs\nr{eom_varphi}--\nr{eom_vr}).
} 
\la{fig:cartoon}
\end{figure}
%

The key for us is the solution of \eq\nr{eom_dr} 
in the presence of $e^{ }_\varphi$. With the given assumptions, 
the right-hand side of \eq\nr{eom_dr} simplifies to 
$ + \Upsilon e^{ }_\varphi$. Replacing $t$ through $H$
as an integration variable, 
the evolution equation can be re-expressed as
\be
 - \partial^{ }_He^{ }_\dr
 + \frac{ 2 ( 1 + w^{ }_{\dr} )}{H} e^{ }_\dr 
 \simeq  \,\frac{ \Upsilon\, \mpl^2 }{4\pi}  
 \;, \quad
 w^{ }_{\dr} \; \equiv \; \frac{p^{ }_{\idr}}{e^{ }_{\idr}}
 \;. \la{eqn_H}
\ee
The general solution is 
\be
 \fbox{$H \gg \Upsilon \gg \Gamma$}: \quad 
 e^{ }_{\varphi} a^3 \simeq \mbox{const}
 \;, \quad
 e^{ }_\dr 
 \; \simeq \;
  C \, H^{2(1+w^{ }_{\idr} )}_{ }
 \; + \; 
 \frac{\mpl^2 \Upsilon H  }{4\pi (1+2 w^{ }_{\idr})} 
 \;, \la{soln}
\ee
where $C$ is an integration constant. 
If $w^{ }_{\dr} > -\tfr{1}{2}$, the latter term dominates
at late times.

An important consequence of \eq\nr{soln} is that we can determine
the phase transition moment, denoted by $H^{ }_*$. 
If $e^{ }_{\dr,*}$ is 
the dark energy density at the phase transition, and omitting the subdominant
term from the right-hand side, \eq\nr{soln} can be inverted into 
\be
 H^{ }_* \gg \Upsilon \gg \Gamma: \quad 
 H^{ }_* \simeq 4 \pi (1+2w^{ }_{*})
 \frac{e^{ }_{\dr, *}}{\mpl^2 \Upsilon}
 \;. \la{H_c}
\ee
Here $w^{ }_*$ is the value of $ w^{ }_{\idr}$ 
at the phase transition point. We argued
above that $p^{ }_{\idr}$ is vanishingly small in the
low-temperature phase, but we keep it in the expressions
here, because then the same formulae can be re-used
in domain~(3), with the substitution $()^{ }_{\idr}\to()^{ }_{\rad}$.

%
\paragraph{Domain~(2).}

Because of $\Upsilon \gg H$, $\varphi$ has equilibrated in this 
domain, and plays little role. Given that $\Gamma \ll H$, 
the right-hand side of \eq\nr{eom_dr} can be omitted, 
and we are faced with 
\be
 \dot{e}^{ }_\dr + 3 H (e^{ }_\dr + p^{ }_\dr )
 \simeq  0
 \;. \la{eom_dr_2} 
\ee
Thus we obtain 
\be
 \fbox{$\Upsilon \gg H  \gg \Gamma$}: \quad 
 e^{ }_{\varphi} \simeq e^{ }_{\varphi,\rmi{eq}}
 \;, \quad
 e^{ }_\dr\, a^{3(1 + w^{ }_\idr)}_{ }
 \simeq \mbox{const}
 \;. \la{soln_2}
\ee
 Let us anticipate that in \se\ref{se:gw}, when we consider 
 gravitational waves generated in a phase transition, 
 only temperatures below the phase transition matter.
 Therefore we can insert  
 $w^{ }_{\idr} \approx 0$ in \eq\nr{soln_2}
 according to our previous assumption.

%
\paragraph{Domain~(3).}

In this domain we can sum together \eqs\nr{eom_dr} and \nr{eom_vr}, 
finding 
\be
 \dot{e}^{ }_{\rad} + 3 H (e^{ }_{\rad} + p^{ }_{\rad} )
 \simeq +\, \Upsilon\, e^{ }_\varphi 
 \;. \la{eom_dr_3}
\ee
The solution is like for \eq\nr{eqn_H}, 
yielding a variant of \eq\nr{soln}, 
\be
 \fbox{$
 \begin{array}{c} 
   \displaystyle H \gg  \Gamma \gg \Upsilon \\
   \displaystyle \Gamma \gg H \gg \Upsilon 
 \end{array} 
 $}
 : \quad 
 e^{ }_{\varphi} a^3 \simeq \mbox{const}
 \;, \quad
 e^{ }_{\rad} 
 \; \simeq \;
 \frac{\mpl^2 \Upsilon H }{4\pi (1+2 w^{ }_{\irad})} 
 \;, \quad
 w^{ }_{\rad} \; \equiv \; \frac{p^{ }_{\irad}}{e^{ }_{\irad}}
 \;. \la{soln_3}
\ee

%
\paragraph{Domain~(4).}

In this domain the solution is like in \eq\nr{soln_2}, except
that the dark and visible sectors have equilibrated, 
\be
 \fbox{$\min\{\Upsilon,\Gamma\} \gg H$}: \quad 
 e^{ }_{\varphi} \simeq e^{ }_{\varphi,\rmi{eq}} 
 \;, \quad
 e^{ }_\rad\, a^{3(1 + w^{ }_\irad)}_{ }
 \; \stackrel{w^{ }_\irad \simeq \frac{1}{3} }{\simeq} \; 
 e^{ }_\rad\, a^{4}_{ }
 \; \simeq \; 
 \mbox{const}
 \;. \la{soln_4}
\ee
If the phase transition takes place in this domain, 
then its gravitational wave signature is not affected
by matter domination, and it redshifts as usual~\cite{old4,ps}.

%
\section{Broad characteristics of the gravitational wave signal}
\la{se:gw}

Considering a phase transition taking place 
in one of the domains mentioned above, 
we wish to work out its gravitational wave signature. 
Here we are concerned with an upper bound, omitting
the complicated hydrodynamics by which it gets formed
(cf.,\ e.g.,\ ref.~\cite{old4}). 
The assumption is that immediately after the transition, gravitational waves
carry the energy density $ e^{ }_{\gw,*} $. The first task
is to estimate which energy fraction this corresponds to today.

We focus on gravitational waves whose wavelength was 
within the horizon at the time of their formation.\footnote{%
 The study of phase transitions leading to longer wavelengths has
 also been initiated~\cite{gh}, however as will be seen in 
 \se\ref{se:max} (cf.\ \fig\ref{fig:strain}), such cases do not 
 extend to the frequencies that are of most interest to us. 
 } 
Their energy density scales with expansion like radiation. 
The relation to the critical energy density today, when the
scale factor is $\anow^{ }$, is 
\be
 \Omega^{ }_{\gw,\now} 
 \; \equiv \;
  \frac{e^{ }_{\gw,\now}}{e^{ }_\rmi{crit}} 
 \; = \;  
  \frac{e^{ }_{\gw,*}}{e^{ }_\rmi{crit}} 
  \frac{a^4_*}{\anow^4}
 \; = \;  
 \underbrace{
  \frac{e^{ }_{\gw,*}}{e^{ }_{\varphi,*}} }_{\it first}
 \,
 \underbrace{
  \frac{e^{ }_{\varphi,*}}{e^{ }_{\rd}}
  \frac{a^4_*}{a^4_\rd} }_{\it second}
 \,
 \underbrace{ 
  \frac{e^{ }_{\rd}}{e^{ }_\rmi{crit}}
  \frac{a^4_\rd}{\anow^4} }_{\it third}
 \;, \la{Omega_gen}
\ee
where $e^{ }_\rmi{crit}$ is the current critical energy density and
the notation $e^{ }_{\pm}$, $a^{ }_\pm$ 
corresponds to that introduced around \eq\nr{H_dd_rd}. 
We have expressed the result as a product of three factors, which
can be estimated as follows, considering the domains
in \fig\ref{fig:cartoon}. 

%
\paragraph{Domain~(1).}

For the {\em first} factor in \eq\nr{Omega_gen}, we parametrize the energy
density released into gravitational radiation by its relation to the
dark sector energy density at that time. Making use of \eq\nr{H_c}, 
this then yields 
\be
 \frac{e^{ }_{\gw,*}}{e^{ }_{\varphi,*}}
 \; = \; 
 \frac{e^{ }_{\gw,*}}{e^{ }_{\dr,*}}
 \frac{e^{ }_{\dr,*}}{e^{ }_{\varphi,*}}
 \; \stackrel{\rmii{\nr{H_c}}}{\simeq} \; 
 \frac{e^{ }_{\gw,*}}{e^{ }_{\dr,*}}
 \frac{\mpl^2 \Upsilon H^{ }_* }{4\pi(1+2w^{ }_*)e^{ }_{\varphi,*}}
 \; \simeq \; 
 \frac{e^{ }_{\gw,*}}{e^{ }_{\dr,*}}
 \frac{2 \Upsilon}{3(1+2w^{ }_*) H^{ }_* }
 \;. \la{Omega_first}
\ee
For the {\em second} factor in \eq\nr{Omega_gen}, we obtain
\be
  \frac{e^{ }_{\varphi,*}}{e^{ }_{\rd}}
  \frac{a^4_*}{a^4_\rd} 
 \; \simeq \;
  \frac{e^{ }_{\varphi,*} a^4_* }{e^{ }_{\dd} a^4_\dd }
  \frac{e^{ }_{\dd} a^4_\dd }{e^{ }_{\rd} a^4_\rd }
 \simeq 
 \biggl( \frac{e^{ }_{\varphi,\dd}}
                     {e^{ }_{\varphi,*}} \biggr)^{1/3}_{ }
 \biggl( \frac{e^{ }_{\idr,\rd}}
                     {e^{ }_{\idr,\dd}}  \biggr)^{1/3}_{ }
 \;\stackrel{\, e^{ }_{\idr,+} \simeq e^{ }_{\varphi,+} \,}{\simeq}\;
 \biggl( \frac{e^{ }_{\idr,\rd}}
                     {e^{ }_{\varphi,*}} \biggr)^{1/3}_{ }
 \simeq 
 \biggl( \frac{H^{ }_\rd}
                     {H^{ }_*}           \biggr)^{2/3}_{ }
 \;.
 \la{Omega_second} 
\ee
For the {\em third} factor, we express $a$ via  
$s^{ }_\rad a_\rad^3 \simeq$ constant, resulting in 
\be
  \frac{e^{ }_{\rd}}{e^{ }_\rmi{crit}}
  \frac{a^4_\rd}{\anow^4}
 \simeq 
  \frac{e^{ }_{\rd}}{e^{ }_\rmi{crit}}
           \frac{\Tnow^4}{T^{4}_\rd}
  \biggl( \frac{ \snow^{ }/ \Tnow^3} 
               { s^{ }_\rd / T^3_\rd} \biggr)^{4/3}_{ }
 \!\!\! = \; 
 \frac{e^{ }_{\gamma,\now}}{e^{ }_\rmi{crit}}
 \frac{e^{ }_{\rd} / T_\rd^4}{e^{ }_{\gamma,\now} / \Tnow^4}
 \biggl( \frac{ g^{ }_{s,\now}} 
              { g^{ }_{s,\rd} } \biggr)^{4/3}_{ }   
\!\!\! = 
 \underbrace{ 
 \frac{g_{s,\now}^{4/3}}{g^{ }_{\gamma,\now}}
 \frac{e^{ }_{\gamma,\now}}{e^{ }_\rmi{crit}}
 \frac{1}{10^{2/3}_{ }}
  }_{ \approx\, 1.65\times 10^{-5}_{ }/ h^{2}_{ } }
 \frac{g^{ }_{e,\rd}}{g^{ }_{s,\rd}} 
 \biggl( 
  \frac{100}{g^{ }_{s,\rd}}
 \biggr)^{1/3}_{ }
 \;, \la{Omega_third}
\ee
where we parametrized the thermal 
energy and entropy densities as
$e^{ }_\rad = g^{ }_{e} \pi^2  T^4 / 30$, 
$s^{ }_\rad = g^{ }_{s} 2 \pi^2 T^3/45$.
For the numerical value, we have employed 
$
 e^{ }_{\gamma,\now} / e^{ }_\rmi{crit} 
 = 2.473\times 10^{-5}_{ } / h^{2}_{ }
$, 
where~$h$ is the reduced Hubble rate, 
as well as $g^{ }_{s,\now} \simeq 3.92$, 
which is related to though not studied
as much as the parameter $N^{ }_\rmi{eff}$ that 
captures the energy density
of the universe after neutrino decoupling~
(cf.\ ref.~\cite[sec.~3.1]{eos2} for a discussion). 
Multiplying together \eqs\nr{Omega_first}--\nr{Omega_third}, finally yields
\be
 h^2\, \Omega^{ }_{\gw,\now} 
 \stackrel{\rm domains\,(1),(3)  }{\simeq}
 1.65\times 10^{-5}_{ }\,
  \frac{g^{ }_{e,\rd}}{g^{ }_{s,\rd}} 
 \biggl( 
  \frac{100}{g^{ }_{s,\rd}}
 \biggr)^{1/3}_{ }
 \biggl(
   \frac{\Upsilon}{H^{ }_*} 
 \biggr)^{ }_{ }
 \biggl(
   \frac{H^{ }_-}{H^{ }_*} 
 \biggr)^{2/3}_{ }
 \frac{2}{3(1+2w^{ }_*)}
  \frac{e^{ }_{\gw,*}}{e^{ }_{\rad,*}}
 \;. \la{Omega_final}
\ee
In the last step we have replaced 
$e^{ }_{\dr,*} \to e^{ }_{\rad,*}$ in domain~(1), 
using the fact that $e^{ }_{\vr,*} \ll e^{ }_{\dr,*}$ there, 
because then the result is also valid in domain~(3). 

%
\paragraph{Domain~(2).}

In this domain, we replace $e^{ }_{\varphi,*}\to e^{ }_{\dr,*}$ 
in \eq\nr{Omega_gen}, because 
$
 e^{ }_{\varphi} \approx e^{ }_{\varphi,\rmi{eq}} \ll e^{ }_{\dr}
$. 
For the {\em second} factor after this modification, 
we still find the same result
as in \eq\nr{Omega_second}, 
\be
 \frac{e^{ }_{\idr,*}}{e^{ }_{\rd}}
 \frac{a^4_*}{a^4_\rd} 
  \;\simeq\;
 \biggl( \frac{e^{ }_{\idr,\rd}}
                     {e^{ }_{\idr,*}} \biggr)^{1/3}_{ }
 \simeq 
 \biggl( \frac{H^{ }_\rd}
                     {H^{ }_*}           \biggr)^{2/3}_{ }
 \;. \la{Omega_second_2}
\ee
The final result becomes 
\be
 h^2\, \Omega^{ }_{\gw,\now} 
 \stackrel{\rm domain\,(2) }{\simeq}
 1.65\times 10^{-5}_{ }\,
  \frac{g^{ }_{e,\rd}}{g^{ }_{s,\rd}} 
 \biggl( 
  \frac{100}{g^{ }_{s,\rd}}
 \biggr)^{1/3}_{ }
 \biggl(
   \frac{H^{ }_-}{H^{ }_*} 
 \biggr)^{2/3}_{ }
  \frac{e^{ }_{\gw,*}}{e^{ }_{\rad,*}}
 \;. \la{Omega_final_2}
\ee
Eq.~\nr{Omega_final} goes over into \nr{Omega_final_2} at the moment
$
 H^{ }_* = 2 \Upsilon /[3(1 + 2w^{ }_*)] \simeq \Upsilon
$.

%
\paragraph{Domain~(3).}

Though intermediate steps differ, 
we recover \eq\nr{Omega_final}. 

%
\paragraph{Domain~(4).}

If the transition takes place in the usual thermal epoch, 
then \eq\nr{Omega_second} is to be omitted, with the result originating
from \eq\nr{Omega_third} evaluated at the transition point, 
\be
 h^2\, \Omega^{ }_{\gw,\now} 
 \stackrel{\rm domain\,(4) }{\simeq}
 1.65\times 10^{-5}_{ }\,
  \frac{g^{ }_{e,*}}{g^{ }_{s,*}} 
 \biggl( 
  \frac{100}{g^{ }_{s,*}}
 \biggr)^{1/3}_{ }
  \frac{e^{ }_{\gw,*}}{e^{ }_{\rad,*}}
 \;. \la{Omega_standard}
\ee

%
\begin{figure}[t]

\centerline{%
     \epsfysize=8.2cm\epsfbox{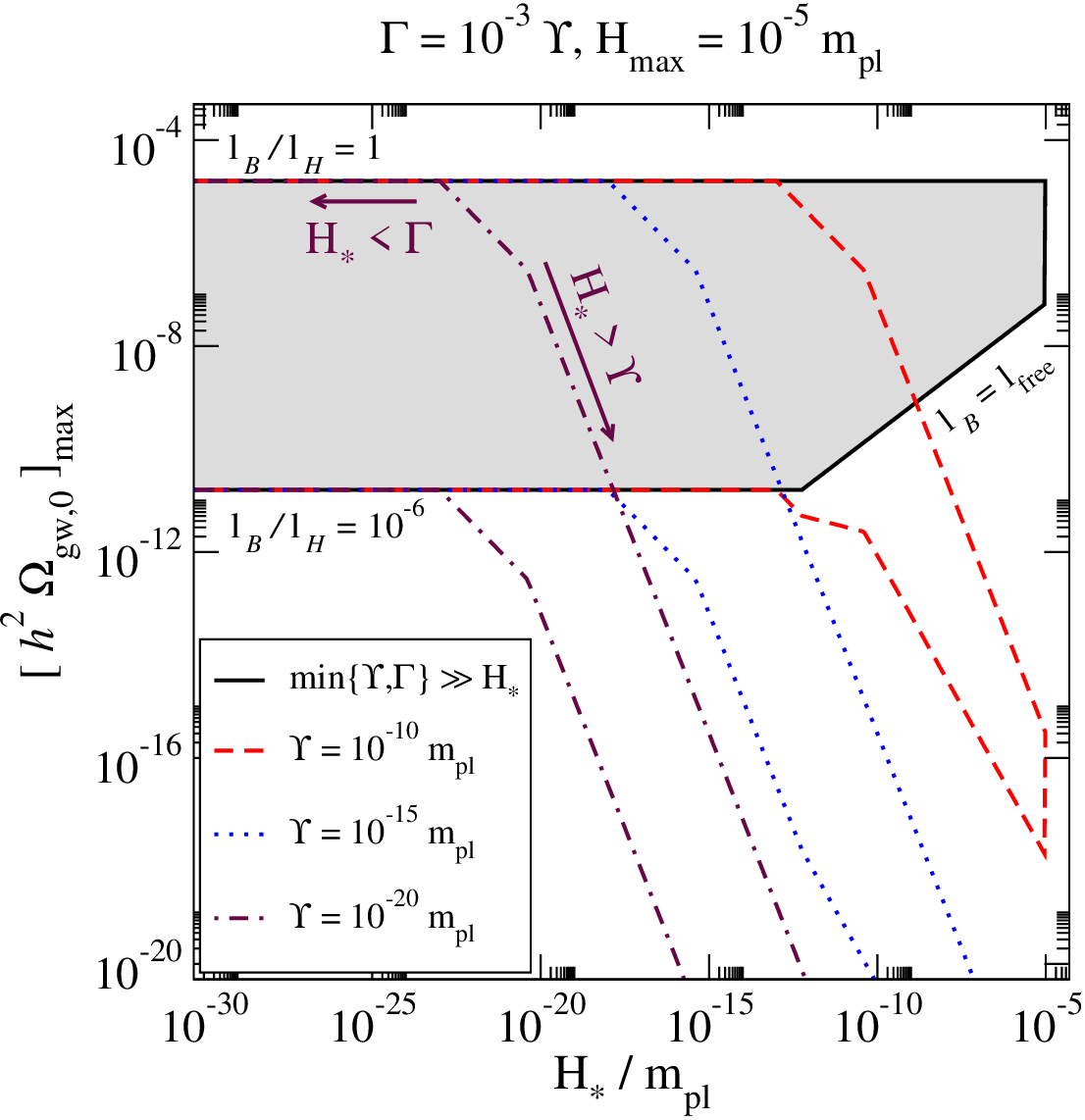}
  ~~~\epsfysize=8.2cm\epsfbox{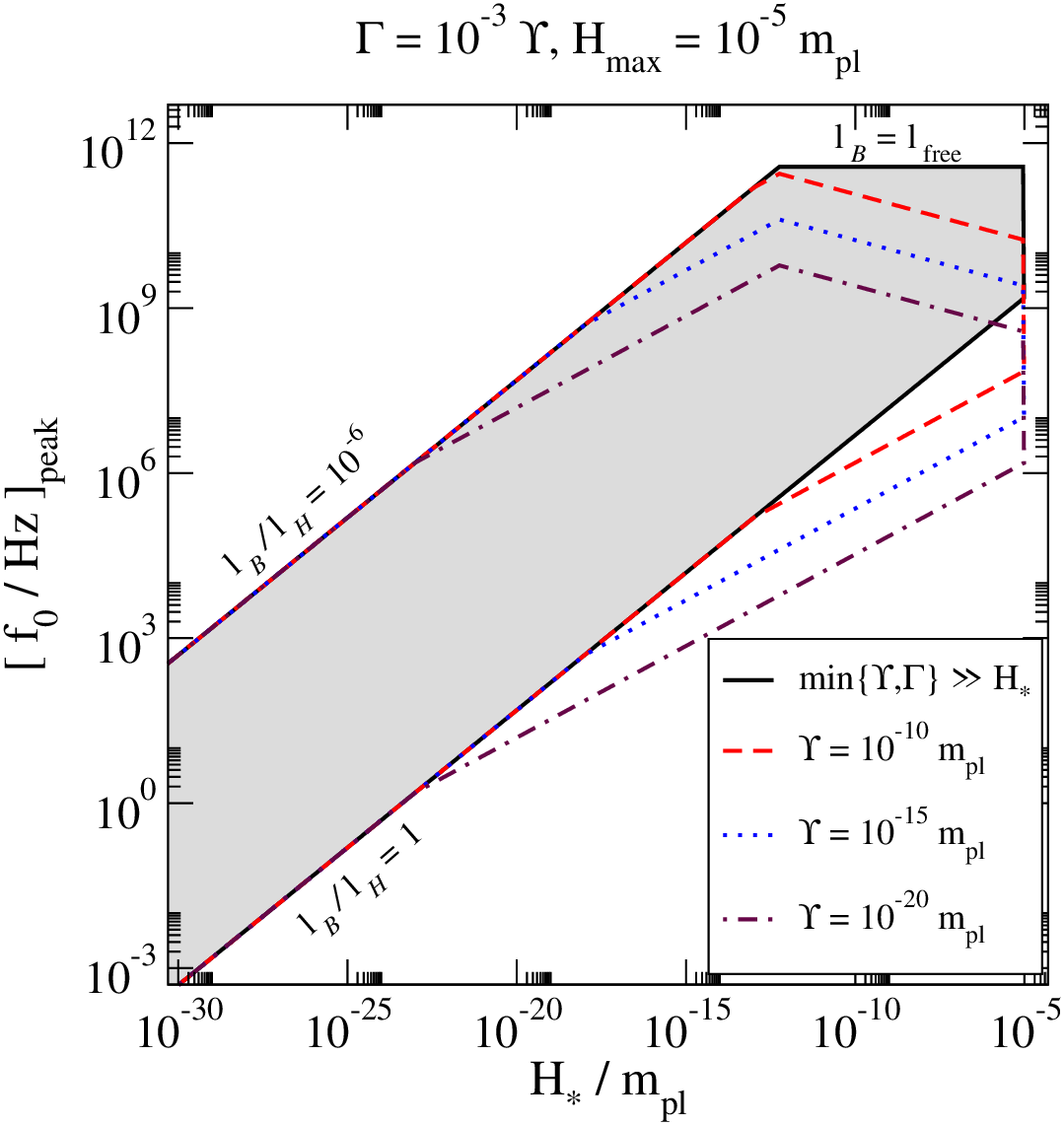}
}

\vspace*{1mm}

\caption[a]{\small 
 Left: the maximal  gravitational
 energy fraction, from \eqs\nr{Omega_final}, 
 \nr{Omega_final_2} and \nr{Omega_standard},
 obtained with 
 $
 {e^{ }_{\igw,*}}/{e^{ }_{\irad,*}} \to
 (\ell^{ }_\B/\ell^{ }_\H)\theta(\ell^{ }_\B - \ell^{ }_\rmii{free})
 $, 
 where 
 $
 \ell^{ }_\B/\ell^{ }_\H
 \in (10^{-6}_{ },1)  
 $. 
 The grey band illustrates the region spanned by variations of 
 $
 \ell^{ }_\B/\ell^{ }_\H
 $, 
 and the dash-dotted (coloured) lines indicate how the boundary
 of the grey band gets modified if the equilibration rates 
 are small, $\min\{\Upsilon,\Gamma\} < H^{ }_*$. 
 We have here chosen $\Gamma = 10^{-3}_{ }\Upsilon$
 for illustration. 
 Only large equilibration rates,
 $\min\{\Upsilon,\Gamma\} \gg H^{ }_*$, permit for
 a sizable effect, otherwise the signal gets diluted. 
 Right: the peak frequency, from 
 \eqs\nr{f0_combine} and \nr{f0_standard}. 
} 
\la{fig:gw}
\end{figure}
%

%
\paragraph{Role of mean free path.}

If microscopic information about the hydrodynamics is inserted,
we can be more precise about the factor 
$
 {e^{ }_{\gw,*}} / {e^{ }_{\rad,*}} 
$
in \eqs\nr{Omega_final}, \nr{Omega_final_2} and \nr{Omega_standard}. 
In particular, since gravitational waves are tensor
excitations, their production rate is proportional
to a bubble length scale breaking translational invariance, 
which we denote in the following by $\ell^{ }_\B$. In fact, 
a quadrupole moment requires a quadratic dependence on 
$\ell^{ }_\B$, but this could be partly compensated for by a long duration
of a process~(cf.,\ e.g.,\ ref.~\cite{simu}). 
Furthermore, if $\ell^{ }_\B$ goes towards zero, 
the production rate does not vanish, but is then taken over 
by that from thermal fluctuations~\cite{gravity_qualitative}. In the so-called
hydrodynamic regime, the fluctuation rate is proportional
to the shear viscosity, which in turn is proportional to 
the mean free path, $\ell^{ }_\rmi{free}$. 
We treat the fluctuation contribution as 
a separate source~(cf.\ \se\ref{se:max}).
To get a conservative upper bound for the phase transition 
contribution, we set 
$
 {e^{ }_{\gw,*}} / {e^{ }_{\rad,*}} 
 \to (\ell^{ }_\B / \ell^{ }_\H) 
 \theta(\ell^{ }_\B - \ell^{ }_\rmi{free})
$, 
where $\ell^{ }_\H \equiv H_*^{-1}$ is the Hubble radius. 
The value of $\ell^{ }_\B/\ell^{ }_\H$ is strongly model dependent, 
so we vary it in the range $1 ... 10^{-6}_{ }$, indicating
the variation as an error band. 


We still need to estimate the numerical value of $\ell^{ }_\rmi{free}$.
If $\alpha \lsim 1$ is 
a characteristic coupling, then the mean free path is 
$\sim 1/ (\alpha^2 T)$, however we would not like to 
make assumptions about the magnitude of the coupling. Therefore 
we set 
\be
 \ell^{ }_\rmi{free} 
 \; \simeq \; 
 \frac{1}{\pi T^{ }_*} 
 \;. \la{constraint}
\ee
The most conservative estimate is obtained
when $\ell^{ }_\rmi{free}$ is smallest, or $T^{ }_*$ is
highest. This is the case when the dark sector energy
density saturates 
$H^{ }_*$, i.e.\ when $\Upsilon > H^{ }_*$.
The results obtained after determining $\ell^{ }_\rmi{free}$
with this recipe are 
shown in \fig\ref{fig:gw}(left).\footnote{%
   We have set $g^{ }_e \simeq g^{ }_s\simeq 106.75$, 
   $w\simeq 1/3$, 
   but these choices have no qualitative effect.
 }

%
\paragraph{Peak frequency.}

The ratio $\ell^{ }_\B / \ell^{ }_\H$ plays an important role
also for the peak frequency of the gravitational wave spectrum.
As we push towards
high temperatures, with correspondingly small values of $\ell^{ }_\H$, 
we must make sure that we do not 
underestimate~$\ell^{ }_\B$, i.e.\ we must maintain
$
 \ell^{ }_\B > \ell^{ }_\rmi{free}. 
$
The peak gravitational wave frequency today, $\fnow^{ }$, is expressed as 
\be
 f^{ }_{\now,\rmi{peak}} 
 \; \equiv \; 
 \frac{a^{ }_*}{\anow} 
 \frac{\theta(\ell^{ }_\B - \ell^{ }_\rmii{free})}{\ell^{ }_\B}
 \; = \; 
 \underbrace{ \frac{a^{ }_*}{a^{ }_\rd} }_{\it first}\;
 \underbrace{ \frac{a^{ }_\rd H^{ }_*}{\anow^{ }} }_{\it second}\;
 \frac{\ell^{ }_\H \,\theta(\ell^{ }_\B - \ell^{ }_\rmii{free}) }{\ell^{ }_\B} 
 \;. \la{f0_def}
\ee
Starting with domain~(1) in \fig\ref{fig:cartoon}, 
the {\em first} factor can 
be approximated as in \eq\nr{Omega_second}, 
yielding
$
 (
   {H^{ }_{-}} / {H^{ }_*} 
 )^{2/3}_{ }
$, 
whereas the {\em second} factor can be expressed like in radiation
domination, 
\ba
 \frac{a^{ }_\rd H^{ }_*}{\anow^{ }} 
 \!\!\! & = & \!\!\! 
 \Tnow^{ } \frac{a^{ }_\rd}{\anow^{ }}
           \frac{H^{ }_*}{\Tnow^{ }}
 \; = \; 
 \Tnow^{ }\biggl( \frac{ \snow^{ }/ \Tnow^3} 
                       { s^{ }_\rd / T^3_\rd} \biggr)^{1/3}_{ }
           \frac{H^{ }_*}{T^{ }_\rd} 
 = 
 \Tnow^{ }\biggl( \frac{ g^{ }_{s,\now}} 
                       { g^{ }_{s,\rd} } \biggr)^{1/3}_{ }
 \biggl( \frac{\pi^2 g^{ }_{e,\rd}}{30} \frac{8\pi}{3\mpl^2}
 \frac{1}{H^2_\rd} \biggr)^{1/4}_{ }\!\! H^{ }_*
 \nn[3mm]
 & \stackrel{ }{\simeq} & 
 \underbrace{ 
             \frac{g_{s,\now}^{1/3}}{10^{1/6}_{ }}
             \biggl( \frac{4\pi^3}{45} \biggr)^{1/4}_{ }
            }_{\approx\, 1.38}
 \, \Tnow^{ }\,
 \biggl(
  \frac{g^{ }_{e,\rd}}{g^{ }_{s,\rd}} 
 \biggr)^{1/4}_{ }
 \biggl( 
  \frac{100}{g^{ }_{s,\rd}}
 \biggr)^{1/12}_{ }
 \biggl(
   \frac{H^{ }_*}{H^{ }_-} 
 \biggr)^{1/2}_{ }
 \biggl(
   \frac{H^{ }_*}{\mpl^{ }} 
 \biggr)^{1/2}_{ }
 \;, \hspace*{5mm} \la{f0_second}
\ea
where
$\Tnow^{ } = 3.57 \times 10^{11}_{ }$\hspace*{0.3mm}Hz. 
Putting the factors together yields
\be
  f^{ }_{\now,\rmi{peak}} 
 \stackrel{\rm domains\,(1),(2),(3) }{\simeq} 
 1.38 \Tnow^{ }\, 
 \biggl(
  \frac{g^{ }_{e,\rd}}{g^{ }_{s,\rd}} 
 \biggr)^{1/4}_{ }
 \biggl( 
  \frac{100}{g^{ }_{s,\rd}}
 \biggr)^{1/12}_{ }
 \biggl(
   \frac{H^{ }_-}{H^{ }_*} 
 \biggr)^{1/6}_{ }
 \biggl(
   \frac{H^{ }_*}{\mpl^{ }} 
 \biggr)^{1/2}_{ }
 \frac{\ell^{ }_\H \,\theta(\ell^{ }_\B - \ell^{ }_\rmii{free}) }{\ell^{ }_\B} 
 \;. \la{f0_combine}
\ee
As indicated, the same formula is obtained in domains~(2) and~(3).
The standard formula for a radiation-dominated epoch
is recovered by omitting the {\em first} factor in \eq\nr{f0_def}, 
and by replacing the 
point between the matter and radiation dominated epochs 
by the phase transition point, 
whereby
\ba
  f^{ }_{\now,\rmi{peak}} 
 & \stackrel{\rm domain\,(4) }{\simeq} & 
 1.38 \Tnow^{ }\, 
 \biggl(
  \frac{g^{ }_{e,*}}{g^{ }_{s,*}} 
 \biggr)^{1/4}_{ }
 \biggl( 
  \frac{100}{g^{ }_{s,*}}
 \biggr)^{1/12}_{ }
 \biggl(
   \frac{H^{ }_*}{\mpl^{ }} 
 \biggr)^{1/2}_{ }
 \frac{\ell^{ }_\H \,\theta(\ell^{ }_\B - \ell^{ }_\rmii{free}) }{\ell^{ }_\B} 
 \;. \hspace*{6mm} \la{f0_standard}
\ea
The results of \eqs\nr{f0_combine} and \nr{f0_standard} 
are plotted in \fig\ref{fig:gw}(right).

%
\section{Maximal gravitational strain from post-inflationary phase transitions}
\la{se:max}

%
\begin{figure}[t]

\centerline{\epsfysize=8.0cm\epsfbox{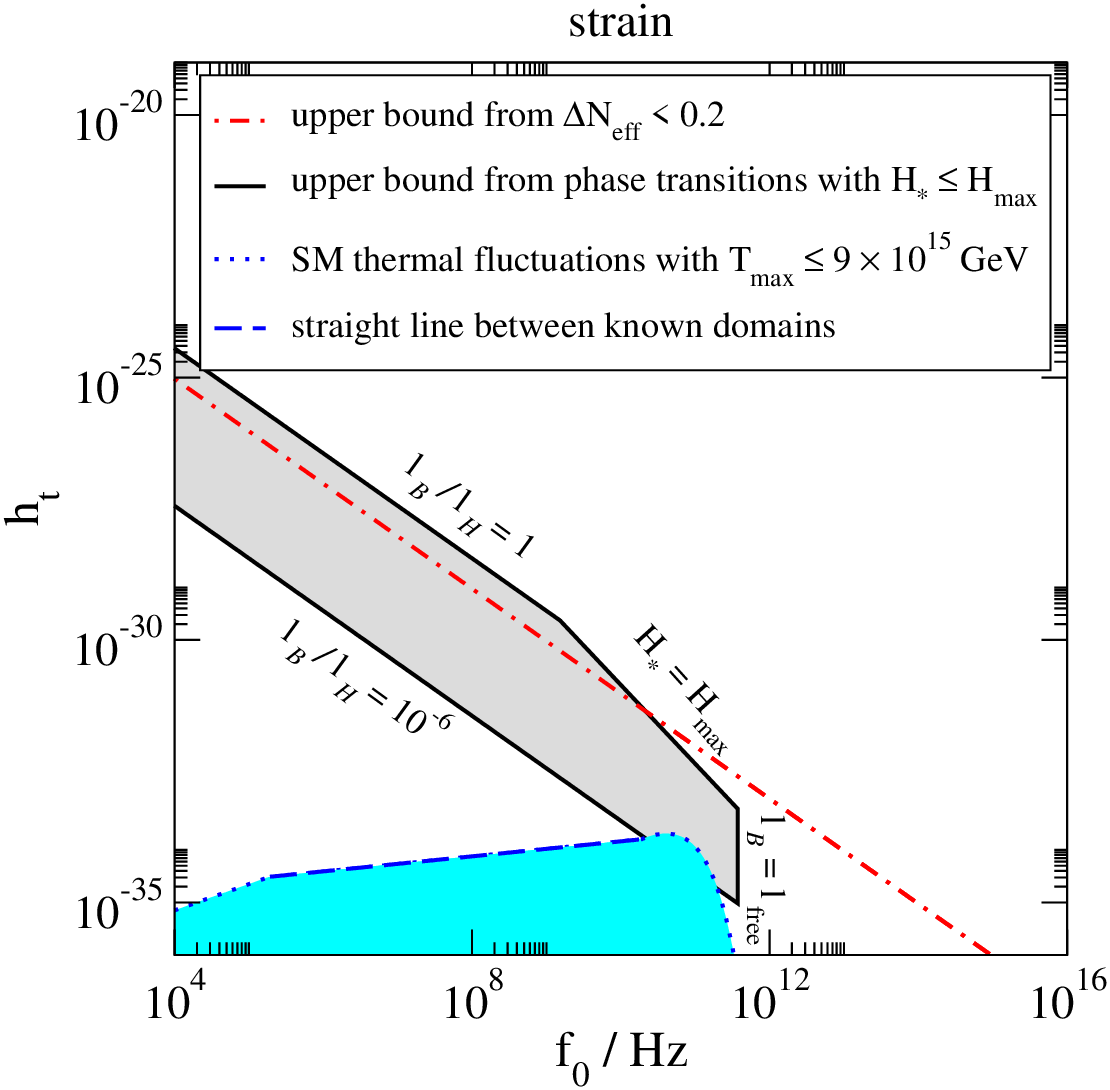}}

\vspace*{1mm}

\caption[a]{\small 
 A combination of the two panels in \fig\ref{fig:gw}
 with $\min\{\Upsilon,\Gamma\} \gg  H^{ }_* $, 
 expressed in terms of the strain $\varh^{ }_\rmi{t}$ from \eq\nr{strain}, 
 compared with the $\Delta N^{ }_\rmii{eff}$ bound from \eq\nr{Neff}, 
 as well as the thermal fluctuation result 
 with $T^{ }_\rmii{max} 
 \le 9 \times 10^{15}_{ }$\hspace*{0.3mm}GeV, 
 corresponding to $H^{ }_\rmii{max} \le 10^{-5}_{ }\mpl^{ }$.
 The last part consists of two known domains, 
 namely an IR tail from hydrodynamic fluctuations
 at small $\fnow^{ }$~\cite{gravity_qualitative}
 and a peak part from elementary particle scatterings at 
 large $\fnow^{ }$~\cite{gravity_lo}; 
 we have connected these with a straight line. 
 The amplitude of the peak part increases in BSM scenarios 
 (cf.\ refs.~\cite{gravity_bsm,reheat} for concrete examples).
} 
\la{fig:strain}
\end{figure}
%

We have seen in \se\ref{se:gw}
that a phase transition during
a matter-dominated epoch, 
present if $\min\{\Upsilon,\Gamma\} < H^{ }_*$, 
leads to a suppressed gravitational wave signal
(cf.\ \eqs\nr{Omega_final} and \nr{Omega_final_2}), 
and reduces the peak frequency of a given transition
(cf.\ \eq\nr{f0_combine}).\footnote{%
 It can be verified that there is suppression also 
 in the gravitational strain, to be introduced below. 
 } 
In the present section we wish to find the maximal possible
signal, at the highest possible peak frequency, 
and therefore consider the regime 
$\min\{\Upsilon,\Gamma\} > H^{ }_*$. In this case 
the transition takes place during radiation domination. 
We represent the results in the same form as 
in \fig{2} of ref.~\cite{uhf}, underlining then also the differences
with respect to this standard reference. 

For quantifying the gravitational wave signal, 
it has become standard to employ
the experimentally meaningful strain, rather than the theoretically
preferred energy density. However, there are a number of possibilities
for its definition.
In terms of \eq\nr{Omega_gen}, the current gravitational energy density
can be expressed as 
$
 e^{ }_{\gw,\now} = 
 \mpl^2 
        \sum_{i,j} 
        \langle\, 
          \dot{\varh}^\rmi{t}_{ij} \dot{\varh}^\rmi{t}_{ij}
        \,\rangle / (32 \pi)
$, 
where $\varh^\rmi{t}_{ij}$ is a metric perturbation in the tensor channel, 
and $\langle ... \rangle$ can be interpreted as a time average. 
The critical energy density is defined as 
$
 e^{ }_\rmi{crit} \equiv 3 \mpl^2 H_\now^2 /(8\pi)
$, 
where $H^{ }_\now$ is the current Hubble rate. 
We go to frequency space
($\partial^{ }_t \to \omega^{ }_\now = 2\pi \fnow^{ }$), and replace
the sum over polarizations by a numerical factor (4, for two polarization
states and an additional factor for the symmetry in $i\leftrightarrow j$),  
whereby $\varh^\rmi{t}_{ij}\to \varh^{ }_\rmi{t}$.
In the absence of spectral information, 
we may effectively assign all the energy density to the peak 
frequency, $f_{\now,\rmi{peak}}^{ }$.
Then the strain can be defined as 
\be
 \Omega^{ }_{\gw,\now} 
 \; \equiv  \; 
 \frac{4\pi^2 f_{\now,\rmi{peak}}^2\, 
 \varh^2_\rmi{t}(f_{\now,\rmi{peak}}^{ })}{3 H_\now^2 }
 \;, \quad 
 \Hnow^{ } =  h\times 3.241\times 10^{-18}_{ } 
 \hspace*{0.3mm}\mbox{Hz}
 \;. \la{strain} 
\ee
Alternatively, if spectral information is available, 
we can parametrize a differential spectrum,
\be
 \frac{ {\rm d} \Omega^{ }_{\gw,\now} }{{\rm d}\ln \fnow^{ }}
 \; \equiv  \; 
 \frac{4\pi^2 f_{\now}^2\, 
 \varh^2_{\rmi{t}\rmii{(alt)}}(f_{\now}^{ })}{3 H_\now^2 }
 \;. \la{strain_2} 
\ee
Writing 
\be
 \frac{ {\rm d} \Omega^{ }_{\gw,\now} }{{\rm d}\ln \fnow^{ }}
 = 
 s(\fnow^{ })
 \, 
 \frac{ {\rm d} \Omega^{ }_{\gw,\now} }{{\rm d}\ln \fnow^{ }}
 \bigg|^{ }_{\fnow^{ } = f^{ }_{\now,\rmii{peak}}}
 \;, \quad
 s(f^{ }_{\now,\rmi{peak}}) \;\equiv\; 1
 \;, 
\ee
we see that 
$
 \varh^2_\rmi{t} (f^{ }_{\now,\rmi{peak}})
 = 
 \varh^2_{\rmi{t}\rmii{(alt)}}(f^{ }_{\now,\rmi{peak}})
 \, 
 \int_{-\infty}^{\infty}
 \! {\rm d} \ln \fnow^{ }
 \, s(\fnow^{ })
$.
Thus the definitions agree at $f^{ }_{\now,\rmi{peak}}$ if 
$
 \int_{-\infty}^{\infty}
 \! {\rm d} \ln \fnow^{ }
 \, s(\fnow^{ })
 \simeq 1.0
$.\footnote{%
 This assumption can be justified 
 up to $\rmO(1)$ for some templates of phase transition spectra.
 }  
Adopting \eq\nr{strain}, 
the results of \fig\ref{fig:gw}
are replotted in \fig\ref{fig:strain}.

We compare the phase transition result 
with two other considerations. The first is 
the parameter $N^{ }_\rmi{eff}$, characterizing the energy density 
carried by additional relativistic species 
at the time of primordial nucleosynthesis. 
If we write the gravitational energy density at that time as
$
 e^{ }_{\gw,\bbn} 
 \equiv 
 \Delta N^{ }_\rmi{eff}\,
 ({7}/{8}) ({4}/{11})^{4/3}_{ } e^{ }_{\gamma,\bbn}
$, 
and redshift until today, then 
\ba
 \Omega^{ }_{\gw,\now} 
  & = & 
  \frac{e^{ }_{\gw,\bbn}}{e^{ }_\rmi{crit}} 
  \frac{a^4_\bbn}{\anow^4}
 \; = \; 
 \Delta N^{ }_\rmi{eff}\,
 \biggl( \frac{s^{ }_\now / \Tnow^3}{s^{ }_\bbn / T^3_\bbn} \biggr)^{4/3}_{ }
 \frac{7}{8}\biggl( \frac{4}{11} \biggr)^{4/3}_{ }
 \biggl( 
 \frac{e^{ }_{\gamma,\bbn} / T^4_\bbn }
      {e^{ }_{\gamma,\now} / \Tnow^4  }
 \biggr)
 \frac{e^{ }_{\gamma,\now} }{ e^{ }_\rmi{crit}  }
 \nn[3mm] 
 & = & 
 \Delta N^{ }_\rmi{eff}\,
 \underbrace{ 
 \biggl( \frac{g^{ }_{s,\now}}{g^{ }_{s,\bbn}} \biggr)^{4/3}_{ }
 \frac{7}{8}\biggl( \frac{4}{11} \biggr)^{4/3}_{ }
 \biggl( 
 \frac{g^{ }_{\gamma,\bbn}  }
      {g^{ }_{\gamma,\now}  }
 \biggr)
 \frac{e^{ }_{\gamma,\now} }{ e^{ }_\rmi{crit}  }
 }_{ \approx\, 5.62 \times 10^{-6}_{ } / h^2}
 \;. \la{Neff}
\ea
We then interpret the upper bound 
$\Delta N^{ }_\rmi{eff} \le 0.2$~\cite{Neff_constraint}
as a bound on 
$
 \varh^{2}_\rmi{t} (f^{ }_{\now,\rmi{peak}})
$
like in \eq\nr{strain}.

The second comparison concerns the 
irreducible background that 
originates from thermal fluctuations.
This contribution is strongly dependent on the maximal 
temperature reached in the early universe. Evaluating
the result by setting $T$ to 
the temperature corresponding to $H^{ }_\rmi{max}$, 
$T^{ }_\rmi{max} \simeq 9\times 10^{15}_{ }$\hspace*{0.3mm}GeV,
the dotted curve in \fig\ref{fig:strain} is obtained, this time 
as an actual spectrum in accordance with \eq\nr{strain_2}.\footnote{%
 The peak part comes from ref.~\cite{gravity_lo},
 which considered physical momenta around $k\sim \pi T$.
 The spectrum is also known in the hydrodynamic 
 domain $k \ll T$~\cite{gravity_qualitative}, 
 or even $k \ll H $~\cite{scan}. 
 In between these domains, no computation exists to date, 
 and our interpolation is simply a straight line. 
 } 
Like for phase transitions (cf.\ \fig\ref{fig:gw}), 
a period of matter domination, 
present if $\min\{\Upsilon,\Gamma\} \ll H^{ }_\rmi{max}$,
would reduce the signal. 

%
\section{Conclusions}
\la{se:concl}

Our main findings can be summarized as follows. 
For phase transitions taking place soon after inflation, 
the gravitational wave signal is suppressed by 
a matter domination epoch, in an analytically
quantifiable manner (cf.\ \eqs\nr{Omega_final} and \nr{Omega_final_2}), 
unless all equilibration
rates are larger than the Hubble rate at the time of the transition. 
In the latter case
(which could be realized for instance in models leading
to a ``strong regime'' of warm inflation, or if the phase
transition takes place after matter domination has ended), 
the phase transition signal can in principle saturate
the $N^{ }_\rmi{eff}$ bound, at $\fnow^{ } < 20$~GHz
(cf.\ \fig\ref{fig:strain}). 
However, whether this actually happens
depends on model-dependent characteristics
of the phase transition.
At $\fnow^{ } > 100$~GHz, in contrast, 
the phase transition signal must merge with the irreducible 
background from thermal fluctuations
(cf.\ \fig\ref{fig:strain}).
The reason is that at distances less than the mean free path, 
``bubbles'' are nothing but regular thermal fluctuations. 
We note that
the latter physics seems to be missing from the considerations 
leading to fig.~2 of ref.~\cite{uhf}.\footnote{%
 In addition, the IR tail from hydrodynamic
 fluctuations appears to have been underestimated there.} 

%
\section*{Acknowledgements}

We thank Simona Procacci for helpful discussions.
The work of H.K.\ was supported by the Research Council of Norway
under the FRIPRO Young Research Talent grant no.\ 335388.

%
\appendix
\renewcommand{\thesection}{Appendix~\Alph{section}}
\renewcommand{\thesubsection}{\Alph{section}.\arabic{subsection}}
\renewcommand{\theequation}{\Alph{section}.\arabic{equation}}

\small{
%

}

\end{document}